# Maximum-Likelihood Decoding of Reed-Solomon Codes is NP-hard

**Venkatesan Guruswami**

Department of Computer Science & Engineering
University of Washington
Box 352350, Seattle, WA 98195, U.S.A.
venkat@cs.washington.edu

**Alexander Vardy**

Department of Electrical and Computer Engineering
Department of Computer Science
Department of Mathematics
University of California, San Diego
9500 Gilman Drive, La Jolla, CA 92093, U.S.A.
vardy@kilimanjaro.ucsd.edu

April 28, 2004

**Abstract**

Maximum-likelihood decoding is one of the central algorithmic problems in coding theory. It has been known for over 25 years that maximum-likelihood decoding of general linear codes is NP-hard. Nevertheless, it was so far unknown whether maximum-likelihood decoding remains hard for *any* specific family of codes with nontrivial algebraic structure. In this paper, we prove that maximum-likelihood decoding is NP-hard for the family of Reed-Solomon codes. We moreover show that maximum-likelihood decoding of Reed-Solomon codes remains hard even with unlimited preprocessing, thereby strengthening a result of Bruck and Naor.

The work of Venkatesan Guruswami was supported by an NSF Career Award. The work of Alexander Vardy was supported in part by the David and Lucile Packard Foundation and by the National Science Foundation.

# 1. Introduction

Maximum-likelihood decoding is one of the central (perhaps, the central) algorithmic problems in coding theory. Berlekamp, McEliece, and van Tilborg [4] showed that this problem is NP-hard for the general class of linear codes. More precisely, the corresponding decision problem can be formally stated as follows.

**Problem:** MAXIMUM-LIKELIHOOD DECODING OF LINEAR CODES (MLD-Linear)
**Instance:** An $m \times n$ matrix $H$ over $\mathbb{F}_q$, a target vector $\underline{s} \in \mathbb{F}_q^m$, and an integer $w > 0$.
**Question:** Is there a vector $\underline{v} \in \mathbb{F}_q^n$ of weight $\leqslant w$, such that $H\underline{v}^t = \underline{s}^t$?

Berlekamp, McEliece, and van Tilborg [4] proved[*] in 1978 that this problem is NP-complete using a reduction from THREE-DIMENSIONAL MATCHING, a well-known NP-complete problem [9, p. 50]. Since 1978, the complexity of maximum-likelihood decoding of general linear codes has been extensively studied. Bruck and Naor [5] and Lobstein [16] showed in 1990 that the problem remains hard even if the code is known in advance, and can be preprocessed for as long as desired in order to devise a decoding algorithm. Arora, Babai, Stern, and Sweedyk [1] proved that MLD-Linear is NP-hard to approximate within any constant factor. Downey, Fellows, Vardy, and Whittle [7] proved that MLD-Linear remains hard even if the parameter $w$ is a constant — it is not fixed-parameter tractable unless $\text{FPT} = W[1]$. Recently, the complexity of approximating MLD-Linear with unlimited preprocessing was studied by Feige and Micciancio [8] and by Regev [19] — this work strengthens the results of both [5, 16] and [1] by showing that MLD-Linear is NP-hard to approximate within a factor of $3 - \varepsilon$ for any $\epsilon > 0$, even if unlimited preprocessing is allowed.

While the papers surveyed in the foregoing paragraph constitute a significant body of work, all these papers deal with the general class of linear codes. This leads to a somewhat incongruous situation. On one hand, there is no nontrivial useful family of codes for which a polynomial-time maximum-likelihood decoding algorithm is known (such a result would, in fact, be regarded a breakthrough). On the other hand, the specific codes used in the reductions of [1, 4, 5, 7, 8, 16, 19] are unnatural, and the problem of showing NP-hardness of maximum-likelihood decoding for *any* useful class of codes with nontrivial algebraic structure remains open, despite repeated calls for its resolution. For example, the survey of algorithmic complexity in coding theory [22] says:

> Although we have, by now, accumulated a considerable amount of results on the hardness of MAXIMUM-LIKELIHOOD DECODING, the broad worst-case nature of these results is still somewhat unsatisfactory. [...] Thus it would be worthwhile to establish the hardness of MAXIMUM-LIKELIHOOD DECODING in the average sense, or for more narrow classes of codes.

The first step along these lines was taken by Alexander Barg [2, Theorem 4], who showed that maximum-likelihood decoding is NP-hard for the class of product (or concatenated) codes, namely codes of type $\mathcal{C} = \mathcal{A} \otimes \mathcal{B}$, where $\mathcal{A}$ and $\mathcal{B}$ are linear codes over $\mathbb{F}_q$. Barg writes in [2] that this result is

> ... the first statement about the decoding complexity of a somewhat more restricted class of codes than just the "general linear codes."

---

[*]Note that MAXIMUM-LIKELIHOOD DECODING OF LINEAR CODES is NP-complete over all finite fields $\mathbb{F}_q$. Berlekamp, McEliece, and van Tilborg [4] only proved this result for the special case $q = 2$. The easy extension to arbitrary prime powers can be found, for instance, in [2, Proposition 2].

Observe, however, that the code $\mathcal{C} = \mathcal{A} \otimes \mathcal{B}$ does not have any algebraic structure unless $\mathcal{A}$ and $\mathcal{B}$ are further restricted in some manner. Furthermore, it is intuitively clear that the decoding problem for this code cannot be much simpler than the decoding problem for its factors $\mathcal{A}$ and $\mathcal{B}$, which are, again, general linear codes.

In this paper, we prove that maximum-likelihood decoding is NP-hard *for the family of Reed-Solomon codes*. Let $q = 2^m$ and let $\mathbb{F}_q[X]$ denote the ring of univariate polynomials over $\mathbb{F}_q$. Reed-Solomon codes are obtained by evaluating certain subspaces of $\mathbb{F}_q[X]$ in a set of points $\mathcal{D} = \{x_1, x_2, \ldots, x_n\}$ which is a subset of $\mathbb{F}_q$. Specifically, a Reed-Solomon code $\mathbb{C}_q(\mathcal{D}, k)$ of length $n$ and dimension $k$ over $\mathbb{F}_q$ is defined as follows:

$$\mathbb{C}_q(\mathcal{D}, k) \stackrel{\text{def}}{=} \left\{ (f(x_1), \ldots, f(x_n)) \ : \ x_1, \ldots, x_n \in \mathcal{D}, \ f(X) \in \mathbb{F}_q[X], \ \deg f(X) < k \right\}$$

Thus a Reed-Solomon code is completely specified in terms of its evaluation set $\mathcal{D}$ and its dimension $k$. As in [4], we assume that if a codeword of $\mathbb{C}_q(\mathcal{D}, k)$ is transmitted and the vector $\underline{y} \in \mathbb{F}_q^n$ is received, the maximum-likelihood decoding task consists of computing a codeword $\underline{c} \in \mathbb{C}_q(\mathcal{D}, k)$ that minimizes $d(\underline{c}, \underline{y})$, where $d(\cdot, \cdot)$ denotes the Hamming distance. The corresponding decision problem can be formally stated as follows.

**Problem:** MAXIMUM-LIKELIHOOD DECODING OF REED-SOLOMON CODES
**Instance:** An integer $m > 0$, a set $\mathcal{D} = \{x_1, x_2, \ldots, x_n\}$ consisting of $n$ distinct elements of $\mathbb{F}_{2^m}$, an integer $k > 0$, a target vector $\underline{y} \in \mathbb{F}_{2^m}^n$, and an integer $w > 0$.
**Question:** Is there a codeword $\underline{c} \in \mathbb{C}_{2^m}(\mathcal{D}, k)$ such that $d(\underline{c}, \underline{y}) \leq w$?

We will refer to this problem[*] as MLD-RS for short. Our main result herein is that MLD-RS is NP-complete. Note that the formulation of MLD-RS is restricted to Reed-Solomon codes over a field of characteristic 2. However, our proof easily extends to Reed-Solomon codes over arbitrary fields: we use fields of characteristic 2 for notational convenience only. The key idea in the proof is a re-interpretation of the result that was derived in [23, Lemma 1] in order to establish NP-hardness of computing the minimum distance of a linear code.

It is particularly interesting that the only nontrivial family of codes for which we can now prove that maximum-likelihood decoding is NP-hard is the family of Reed-Solomon codes. Decoding of Reed-Solomon codes is a well-studied problem with a long history. There are well-known polynomial-time algorithms that decode Reed-Solomon codes up to half their minimum distance [3, 10, 18], and also well beyond half the minimum distance [12, 21].

---
[*]In the definition of MLD-RS, the field elements of $\mathbb{F}_{2^m}$ are assumed to be represented by $m$-bit vectors. Therefore the input size of an instance of MLD-RS is polynomial in $n$ and $m$.

Nevertheless, all these algorithms fall in the general framework of bounded-distance decoders [22]. Our result shows that assuming a bound on the number of correctable errors, as these algorithms do, is necessary, since maximum-likelihood decoding is NP-hard.

In terms of work with related results, Goldreich, Rubinfeld, and Sudan [11] considered a problem similar to MLD-RS in the context of general polynomial reconstruction problems. Thus it is shown in [11, Section 6.1] that given $n$ pairs $(x_1, y_1), (x_2, y_2), \ldots, (x_n, y_n)$ of elements from a large field, determining if a degree $k$ polynomial passes through at least $k + 2$ of them is NP-hard. However, this formulation does *not* include the essential restriction that the evaluation points $x_1, x_2, \ldots, x_n$ are all distinct (in fact, the proof of [11] crucially exploits the fact that $x_i = x_j$ for some $i \neq j$), and therefore does not yield any hardness results for Reed-Solomon decoding. We show that a problem very similar to the one considered in [11] remains NP-hard when the evaluation points $x_1, x_2, \ldots, x_n$ *are* distinct. Thus our result can be viewed as resolving one of the main questions left open by [11].

The proof of our main result (Theorem 5) is presented in the next section. In Section 3, we further strengthen this result by showing that maximum-likelihood decoding of Reed-Solomon codes remains hard even if unlimited preprocessing is allowed, and only the received vector $y$ is part of the input. This is a well-motivated scenario, since the code (namely, the evaluation set $\mathcal{D}$ and the dimension $k$) is usually known in advance. Thus one-time preprocessing, even if computationally expensive, would be attractive if it leads to efficient decoding. We prove in Section 3 (assuming NP does not have polynomial-size circuits) that for some Reed-Solomon codes no such preprocessing procedure can exist. This strengthens the main result of Bruck and Naor [5] in the same way that Theorem 5 strengthens the main result of Berlekamp, McEliece, and van Tilborg [4]. We conclude the paper in Section 4 with a brief discussion, pointing out several simple corollaries of Theorem 5 and suggesting a number of interesting open problems related to our results.

## 2. MLD-RS is NP-complete

As in Berlekamp, McEliece, and van Tilborg [4], we reduce from THREE-DIMENSIONAL MATCHING. Let $\mathcal{U} = \{1, 2, \ldots, t\}$ and let $\mathcal{T}$ be a set of ordered triples over $\mathcal{U}$, that is $\mathcal{T} \subseteq \mathcal{U} \times \mathcal{U} \times \mathcal{U}$. A subset $\mathcal{S}$ of $\mathcal{T}$ is called a **matching** if $|\mathcal{S}| = t$ and every two triples in $\mathcal{S}$ differ in all three positions. As shown by Karp in his seminal paper [13] back in 1972, the following decision problem is NP-complete.

**Problem:** THREE-DIMENSIONAL MATCHING
**Instance:** A set of ordered triples $\mathcal{T} \subseteq \{1, 2, \ldots, t\} \times \{1, 2, \ldots, t\} \times \{1, 2, \ldots, t\}$.
**Question:** Is there a matching in $\mathcal{T}$, namely a subset $\mathcal{S} \subseteq \mathcal{T}$ consisting of exactly $t$ triples such that $d(\underline{s}, \underline{s}') = 3$ for all distinct $\underline{s}, \underline{s}' \in \mathcal{S}$?

We shall write an instance of THREE-DIMENSIONAL MATCHING as $\{t, \mathcal{T}\}$. We henceforth assume w.l.o.g. that $|\mathcal{T}| > t + 1$ (otherwise, the problem is trivially solvable in poly-

nomial time). The following deterministic procedure converts any such instance $\{t,\mathcal{T}\}$ into an instance $\{m,\mathcal{D},k,w,\underline{y}\}$ of MLD-RS.

**A. COMPUTING THE INTEGER PARAMETERS:** Set $m = 3t$, $k = |\mathcal{T}| - (t+1)$, and $w = t$. Let $n = |\mathcal{T}|$.

**B. COMPUTING THE EVALUATION SET:** Let $q = 2^m$. First, construct the finite field $\mathbb{F}_q$ — that is, generate a primitive irreducible (over $\mathbb{F}_2$) binary polynomial of degree $m$ which defines addition and multiplication in $\mathbb{F}_q$. Let $\alpha$ denote a root of this polynomial. Then $\alpha$ is a primitive element of $\mathbb{F}_q$ and the set $\{1, \alpha, \alpha^2, \ldots, \alpha^{m-1}\}$ is a basis for $\mathbb{F}_q$ over $\mathbb{F}_2$. Now, convert each triple $(a,b,c) \in \mathcal{T}$ into a nonzero element of $\mathbb{F}_q$ as follows:

$$(a,b,c) \mapsto x = \alpha^{a-1} + \alpha^{t+b-1} + \alpha^{2t+c-1} \qquad (1)$$

This produces $n = |\mathcal{T}|$ distinct nonzero elements $x_1, x_2, \ldots, x_n \in \mathbb{F}_q$. Set the evaluation set $\mathcal{D}$ to $\{x_1, x_2, \ldots, x_n\}$.

**C. COMPUTING THE TARGET VECTOR:** Compute $\gamma = 1 + \alpha + \cdots + \alpha^{m-1}$ in $\mathbb{F}_q$. Thus $\gamma$ is the element of $\mathbb{F}_q$ that corresponds to the binary $m$-tuple $(1,1,\ldots,1)$ in the chosen basis. Now, for each $j = 1, 2, \ldots, w+1$, compute

$$z_j \stackrel{\text{def}}{=} \frac{\gamma - \sum_{\substack{i=1 \\ i \neq j}}^{w+1} x_i}{\prod_{\substack{i=1 \\ i \neq j}}^{w+1} (x_j - x_i)} \qquad \text{and} \qquad \varphi_j \stackrel{\text{def}}{=} \prod_{\beta \in \mathbb{F}_q \setminus \mathcal{D}} (x_j - \beta) \qquad (2)$$

Note that $\varphi_1, \varphi_2, \ldots, \varphi_{w+1}$ are all nonzero by definition. Set the target vector $\underline{y} = (y_1, y_2, \ldots, y_n)$ to $(z_1/\varphi_1, z_2/\varphi_2, \ldots, z_{w+1}/\varphi_{w+1}, 0, 0, \ldots, 0)$. In other words: $y_j = z_j/\varphi_j$ for $j = 1, 2, \ldots, w+1$, and $y_j = 0$ otherwise.

We will refer to the foregoing computation as the 3-DM/MLD-RS *conversion procedure*. It is not immediately clear that this procedure runs in polynomial time (note that the conversion procedure has to run in time which is polynomial in the size of the THREE-DIMENSIONAL MATCHING instance $\{t, \mathcal{T}\}$, and therefore in time that is polynomial in the logarithm of the field size). This fact is, therefore, established in the following lemma.

**Lemma 1.** *The 3-DM/MLD-RS conversion procedure runs in time and space that are bounded by a polynomial in the size of the instance $\{t, \mathcal{T}\}$.*

*Proof.* Step A is trivial. The only thing that is not immediately obvious in Step B is whether a primitive irreducible binary polynomial of degree $m = 3t$ can be generated in deterministic polynomial time. However, Shoup [20] provides a deterministic algorithm for this purpose, whose complexity is $O(m^5)$ operations in $\mathbb{F}_2$. Clearly, $\gamma$ and $z_1, z_2, \ldots, z_{w+1}$ in Step C can be computed in polynomial time and space. However, it is not clear whether this is also true for $\varphi_1, \varphi_2, \ldots, \varphi_{w+1}$. Indeed, a straightforward evaluation of the expression for $\varphi_j$ in (2) takes $2^m - n$ additions and multiplications in $\mathbb{F}_q$. Thus we now show how to compute $\varphi_j$ in polynomial time. Define the polynomials

$$M(X) \stackrel{\text{def}}{=} \prod_{\beta \in \mathbb{F}_q} (X - \beta) = X^q - X \tag{3}$$

$$D(X) \stackrel{\text{def}}{=} \prod_{\beta \in \mathcal{D}} (X - \beta) = \sum_{i=0}^{n} d_i X^i \tag{4}$$

and let $G(X)$ denote the rational function $M(X)/D(X)$. Then $\varphi_j = G(x_j)$ in view of (2). It is easy to see from (3) and (4) that $x_j$ is a simple root of both $M(X)$ and $D(X)$. Hence

$$G(x_j) = \frac{M'(x_j)}{D'(x_j)}$$

where $M'(X)$ and $D'(X)$ are the first-order Hasse derivatives of $M(X)$ and $D(X)$, respectively. Note that $M'(X) = 1$ in a field of characteristic 2. It follows that

$$\varphi_j = \frac{1}{\sum_{i=0}^{n'} d_{2i+1} x_j^{2i}} \qquad \text{for } j = 1, 2, \ldots, w+1 \tag{5}$$

where $n' = \lfloor (n-1)/2 \rfloor$ and the coefficients $d_0, d_1, \ldots, d_n$ are elementary symmetric functions of $x_1, x_2, \ldots, x_n$. These coefficients can be computed from (4) in time $O(n^2)$. Given $d_0, d_1, \ldots, d_n$, the computation in (5) clearly requires at most $O(wn)$ operations in $\mathbb{F}_q$. ∎

Let $H = [h_{i,j}]$ be the $(w+1) \times n$ matrix over $\mathbb{F}_q$ defined by $h_{i,j} = x_j^{i-1}$ for $j = 1, 2, \ldots, n$ and $i = 1, 2, \ldots, w+1$, where $x_1, x_2, \ldots, x_n$ are given by (1). Explicitly

$$H \stackrel{\text{def}}{=} \begin{bmatrix} 1 & 1 & \cdots & 1 \\ x_1 & x_2 & \cdots & x_n \\ x_1^2 & x_2^2 & \cdots & x_n^2 \\ \vdots & \vdots & & \vdots \\ x_1^w & x_2^w & \cdots & x_n^w \end{bmatrix} \tag{6}$$

The following lemma is a key step in our reduction from THREE-DIMENSIONAL MATCHING to MLD-RS. This lemma owes its general idea to [23, Lemma 1].

**Lemma 2.** *The set $\mathcal{T}$ has a matching if and only if there is a vector $\underline{v} \in \mathbb{F}_q^n$ of weight $\leqslant w$ such that $H \underline{v}^t = (0, 0, \ldots, 0, 1, \gamma)^t$.*

*Proof.* Following Berlekamp, McEliece, and van Tilborg [4], we first construct the $m \times n$ (or $3t \times |\mathcal{T}|$) binary matrix $V$ having the binary representations of $x_1, x_2, \ldots, x_n$ as its columns. As noted in [4], $\mathcal{T}$ has a matching if and only if there is a set of $w = t$ columns of $V$ that add to the all-one vector. The latter condition can be equivalently stated over $\mathbb{F}_q$

as follows: there is a subset $\{x_{i_1}, x_{i_2}, \ldots, x_{i_w}\}$ of $\mathcal{D}$ such that $x_{i_1} + x_{i_2} + \cdots + x_{i_w} = \gamma$. Suppose that $\mathcal{T}$ has a matching, so that such a set $\{x_{i_1}, x_{i_2}, \ldots, x_{i_w}\} \subset \mathcal{D}$ exists, and consider the matrix

$$A = \begin{bmatrix} 1 & 1 & \cdots & 1 & 0 \\ x_{i_1} & x_{i_2} & \cdots & x_{i_w} & 0 \\ x_{i_1}^2 & x_{i_2}^2 & \cdots & x_{i_w}^2 & 0 \\ \vdots & \vdots & & \vdots & \vdots \\ x_{i_1}^{w-2} & x_{i_2}^{w-2} & \cdots & x_{i_w}^{w-2} & 0 \\ x_{i_1}^{w-1} & x_{i_2}^{w-1} & \cdots & x_{i_w}^{w-1} & 1 \\ x_{i_1}^{w} & x_{i_2}^{w} & \cdots & x_{i_w}^{w} & \gamma \end{bmatrix} \qquad (7)$$

It was shown in [23, Lemma 1] that

$$\det A = (\gamma - x_{i_1} - x_{i_2} - \cdots - x_{i_w}) \prod_{1 \leq a < b \leq w} (x_{i_b} - x_{i_a}) \qquad (8)$$

Hence $A$ is singular, so there exists a nonzero vector $\underline{u} = (u_1, u_2, \ldots, u_{w+1}) \in \mathbb{F}_q^{w+1}$ such that $A\underline{u}^t = \mathbf{0}$. We claim that $u_{w+1} \neq 0$. To see this, replace the last column of $A$ by the vector $(1, 1, \ldots, 1)^t$ to obtain the $(w+1) \times (w+1)$ matrix $A'$. If $u_{w+1} = 0$ then $A'\underline{u}^t = \mathbf{0}$, which is a contradiction since $\det A'$ is clearly nonzero (as $x_j \neq 1$ for all $j$ by (1), it is the determinant of a Vandermonde matrix with distinct columns). We can now construct a vector $\underline{v} = (v_1, v_2, \ldots, v_n) \in \mathbb{F}_q^n$ of weight $\leq w$ as follows

$$v_j = \begin{cases} -\dfrac{u_{i_r}}{u_{w+1}} & x_j = x_{i_r} \text{ for some } r \in \{1, 2, \ldots, w\} \\ 0 & \text{otherwise} \end{cases}$$

It should be obvious from (6), (7) and the fact that $A\underline{u}^t = \mathbf{0}$ that $H\underline{v}^t = (0, 0, \ldots, 0, 1, \gamma)^t$. Conversely, assume that there is a vector $v \in \mathbb{F}_q^n$ such that $H\underline{v}^t = (0, 0, \ldots, 0, 1, \gamma)^t$ and $\text{wt}(\underline{v}) \leq w$. Write $\delta = \text{wt}(\underline{v})$ and let $\{i_1, i_2, \ldots, i_\delta\}$ be the set of nonzero positions of $\underline{v}$. Let $\{i_{\delta+1}, i_{\delta+2}, \ldots, i_w\}$ be an arbitrary subset of $\{1, 2, \ldots, n\}$ of size $w - \delta$, that is disjoint from $\{i_1, i_2, \ldots, i_\delta\}$. Then, as in (8), we have

$$0 = \begin{vmatrix} 1 & 1 & \cdots & 1 & 0 \\ x_{i_1} & x_{i_2} & \cdots & x_{i_w} & 0 \\ x_{i_1}^2 & x_{i_2}^2 & \cdots & x_{i_w}^2 & 0 \\ \vdots & \vdots & & \vdots & \vdots \\ x_{i_1}^{w-2} & x_{i_2}^{w-2} & \cdots & x_{i_w}^{w-2} & 0 \\ x_{i_1}^{w-1} & x_{i_2}^{w-1} & \cdots & x_{i_w}^{w-1} & 1 \\ x_{i_1}^{w} & x_{i_2}^{w} & \cdots & x_{i_w}^{w} & \gamma \end{vmatrix} = (\gamma - x_{i_1} - x_{i_2} - \cdots - x_{i_w}) \prod_{1 \leq a < b \leq w} (x_{i_b} - x_{i_a}) \qquad (9)$$

since the fact that $H\underline{v}^t = (0, 0, \ldots, 0, 1, \gamma)^t$ implies that the matrix in (9) is singular. Since $x_1, x_2, \ldots, x_n$ are all distinct, it follows from (9) that $x_{i_1} + x_{i_2} + \cdots + x_{i_w} = \gamma$. This, in turn, implies that there is a matching in $\mathcal{T}$, and we are done. ∎

Recall that $k = |\mathcal{T}| - (t+1) = n - (w+1)$ in our conversion procedure, and let $\mathcal{C}$ be the $(n,k)$ linear code over $\mathbb{F}_q$ having the matrix $H$ in (6) as its parity-check matrix. Further, let $\underline{z} = (z_1, z_2, \ldots, z_{w+1}, 0, 0, \ldots, 0) \in \mathbb{F}_q^n$, where $z_1, z_2, \ldots, z_{w+1}$ are defined by (2).

**Corollary 3.** *The set $\mathcal{T}$ has a matching if and only if the code $\mathcal{C} \stackrel{\text{def}}{=} \{\underline{v} \in \mathbb{F}_q^n : H\underline{v}^t = \mathbf{0}\}$ contains a codeword at Hamming distance $\leqslant w$ from $\underline{z}$.*

*Proof.* In view of Lemma 2, it would suffice to show that the syndrome of $\underline{z}$ with respect to $H$ is $(0, 0, \ldots, 0, 1, \gamma)^t$. Explicitly, we need to prove that

$$H\underline{z}^t = \begin{bmatrix} 1 & 1 & \cdots & 1 \\ x_1 & x_2 & \cdots & x_{w+1} \\ x_1^2 & x_2^2 & \cdots & x_{w+1}^2 \\ \vdots & \vdots & & \vdots \\ x_1^w & x_2^w & \cdots & x_{w+1}^w \end{bmatrix} \begin{bmatrix} z_1 \\ z_2 \\ z_3 \\ \vdots \\ z_{w+1} \end{bmatrix} = \begin{bmatrix} 0 \\ 0 \\ \vdots \\ 0 \\ 1 \\ \gamma \end{bmatrix} \qquad (10)$$

The easiest way to see that the second equality in (10) holds is to regard this as a system of linear equations in the indeterminates $z_1, z_2, \ldots, z_{w+1}$. Let $M$ denote the $(w+1) \times (w+1)$ matrix in (10). Since $M$ is clearly nonsingular, the system admits a unique solution, given by $z_j = \det M_j / \det M$ for $j = 1, 2, \ldots, w+1$, where $M_j$ is the matrix obtained from $M$ by replacing the $j$-th column with $(0, 0, \ldots, 0, 1, \gamma)^t$. Now

$$\det M_j = \pm \left( \gamma - \sum_{\substack{i=1 \\ i \neq j}}^{w+1} x_i \right) \prod_{\substack{1 \leqslant a < b \leqslant w+1 \\ a,b \neq j}} (x_b - x_a) \qquad \text{for } j = 1, 2, \ldots, w+1$$

as in (8), while $\det M$ is the Vandermonde determinant $\prod_{1 \leqslant a < b \leqslant w+1}(x_b - x_a)$. From this, the expression for $z_j$ in (2) easily follows. ∎

The last observation we need is that the code $\mathcal{C}$ defined in Corollary 3 is just a generalized Reed-Solomon code. Specifically, let us extend the definition of $\varphi_1, \varphi_2, \ldots, \varphi_{w+1}$ in (2) to all $j = 1, 2, \ldots, n$ and consider the mappings $\varphi : \mathbb{F}_q^n \to \mathbb{F}_q^n$ and $\varphi^{-1} : \mathbb{F}_q^n \to \mathbb{F}_q^n$ defined by

$$\varphi(u_1, u_2, \ldots, u_n) \stackrel{\text{def}}{=} (\varphi_1 u_1, \varphi_2 u_2, \ldots, \varphi_n u_n)$$

$$\varphi^{-1}(u_1, u_2, \ldots, u_n) \stackrel{\text{def}}{=} (u_1/\varphi_1, u_2/\varphi_2, \ldots, u_n/\varphi_n)$$

Note that $\varphi^{-1}$ is well-defined since $\varphi_1, \varphi_2, \ldots, \varphi_n$ are all nonzero. Also note that both $\varphi$ and $\varphi^{-1}$ are bijections and isometries with respect to the Hamming distance.

**Lemma 4.**
$$\varphi^{-1}(\mathcal{C}) = \mathbb{C}_q(\mathcal{D}, k)$$

*Proof.* We will prove the equivalent statement that $\mathcal{C}$ is the image of $\mathbb{C}_q(\mathcal{D}, k)$ under $\varphi$. Let $G = [g_{i,j}]$ be the $k \times n$ matrix over $\mathbb{F}_q$ defined by $g_{i,j} = x_j^{i-1}$ for all $i = 1, 2, \ldots, k$ and $j = 1, 2, \ldots, n$. It is well known (and obvious) that $G$ is a generator matrix for $\mathbb{C}_q(\mathcal{D}, k)$.

Hence a generator matrix for the image of $\mathbb{C}_q(\mathcal{D},k)$ under $\varphi$ is given by $G' = [g'_{i,j}]$ where $g'_{i,j} = \varphi_j x_j^{i-1}$. It would therefore suffice to prove that $G'$ is a generator matrix for the code $\mathcal{C}$, which is equivalent to the statement that $B = G'H^T$ is the $k \times (w+1)$ all-zero matrix. By definition, a generic entry of $B = [b_{r,s}]$ is given by

$$b_{r,s} = \sum_{j=1}^n g'_{r,j} h_{s,j} = \sum_{j=1}^n \varphi_j x_j^{r-1} x_j^{s-1} = \sum_{j=1}^n \varphi_j x_j^{r+s-2} \qquad (11)$$

for $r = 1,2,\ldots,k$ and $s = 1,2,\ldots,w+1$. Now, let $\mathbb{F}_q^*$ denote the set of nonzero elements in $\mathbb{F}_q$, and define the polynomials

$$\Psi(X) \stackrel{\text{def}}{=} \prod_{\beta \in \mathbb{F}_q^* \setminus \mathcal{D}} (X - \beta) = \sum_{j=0}^{q-n-1} \psi_j X^j \qquad (12)$$

$$\Phi(X) \stackrel{\text{def}}{=} X\Psi(X) = \prod_{\beta \in \mathbb{F}_q \setminus \mathcal{D}} (X - \beta) \qquad (13)$$

By the definition of $\varphi_j$ in (2), we have $\varphi_j = \Phi(x_j) = x_j \Psi(x_j)$ for all $j = 1,2,\ldots,n$. Substituting this in (11), we obtain

$$b_{r,s} = \sum_{j=1}^n x_j \Psi(x_j) x_j^{r+s-2} = \sum_{\beta \in \mathbb{F}_q^*} \Psi(\beta) \beta^{r+s-1} = \sum_{\beta \in \mathbb{F}_q^*} \sum_{j=0}^{q-n-1} \psi_j \beta^j \beta^{r+s-1} \qquad (14)$$

where the second equality follows from the fact that $\Psi(\beta) = 0$ for all $\beta \in \mathbb{F}_q^* \setminus \mathcal{D}$. Finally, interchanging the order of summation in (14), we obtain

$$b_{r,s} = \sum_{j=0}^{q-n-1} \psi_j \sum_{\beta \in \mathbb{F}_q^*} \beta^{j+r+s-1} = \sum_{j=0}^{q-n-1} \psi_j \sum_{i=0}^{q-2} (\alpha^i)^{j+r+s-1} = \sum_{j=0}^{q-n-1} \psi_j \sum_{i=0}^{q-2} \xi^i \qquad (15)$$

where $\alpha$ is a primitive element of $\mathbb{F}_q$ and $\xi = \alpha^{j+r+s-1}$. The last summation in (15) is a geometric series which evaluates to $(\xi^{q-1} - 1)/(\xi - 1) = 0$ provided $\xi \neq 1$. However, since $2 \leqslant r+s \leqslant n$, it is easy to see that we will always have $1 \leqslant j+r+s-1 \leqslant q-2$. Hence $\xi = \alpha^{j+r+s-1} \neq 1$. Thus $b_{r,s} = 0$ for all $r$ and $s$, and the lemma follows. ∎

We are now ready to prove our main result in this paper. Indeed, all that remains to be done to establish that MLD-RS is NP-complete is to combine Lemma 4 with Corollary 3.

**Theorem 5.** *MLD-RS is NP-complete.*

*Proof.* Note that $\underline{y} = \varphi^{-1}(\underline{z})$ in the 3-DM/MLD-RS conversion procedure. Since $\varphi^{-1}$ is an isometry, it follows from Lemma 4 that there is a codeword $\underline{c} \in \mathbb{C}_q(\mathcal{D},k)$ such that $d(\underline{c},\underline{y}) \leqslant w$ if and only if $\mathcal{C}$ contains a codeword at distance $\leqslant w$ from $\underline{z}$. By Corollary 3, this happens iff the set $\mathcal{T}$ has a matching. Hence the 3-DM/MLD-RS conversion procedure is a polynomial transformation from THREE-DIMENSIONAL MATCHING to MLD-RS. ∎

## 3. Hardness of MLD-RS with preprocessing

As noted in [5, 16], the formulation of MLD-RS in the previous two sections might not be the relevant one in practice. In coding practice, the code to be decoded is usually known in advance; moreover, this code remains the same throughout numerous decoding attempts wherein only the target vector $\underline{y}$ changes. Thus it would make sense to assume that the code is known *a priori* and can be preprocessed for a long time (essentially, unlimited time) in order to devise an efficient decoding algorithm.

In the special case of Reed-Solomon codes, the general observation above reduces to the following assumption: the Reed-Solomon code $\mathbb{C}_q(\mathcal{D}, k)$ — namely, the set of evaluation points $\mathcal{D} = \{x_1, x_2, \ldots, x_n\} \subseteq \mathbb{F}_q$ and the dimension $k$ — is known in advance (and can be preprocessed for as long as desired) and only the target vector $\underline{y} \in \mathbb{F}_q^n$ is part of the input. The corresponding decision problem can be formally phrased as follows.

**Problem:** MLD-RS WITH PREPROCESSING
**Instance:** A target vector $\underline{y} \in \mathbb{F}_{2^m}^n$.
**Question:** Is there a codeword $\underline{c} \in \mathbb{C}_{2^m}(\mathcal{D}, k)$ such that $d(\underline{c}, \underline{y}) \leqslant w$?

Observe that the above defines not one problem, but a whole set of problems — one for each realization of $m$, $\mathcal{D}$, $k$, and $w$. We shall henceforth refer to a specific problem in this set as MLD-RSwP$(m, \mathcal{D}, k, w)$. Asking whether a given problem MLD-RSwP$(m, \mathcal{D}, k, w)$ is computationally hard makes no sense, since the size of the input $\underline{y} \in \mathbb{F}_{2^m}^n$ to this problem is at most $mn$ bits, while both $m$ and $n = |\mathcal{D}|$ are fixed. Thus asymptotic complexity questions concerning a specific problem MLD-RSwP$(m, \mathcal{D}, k, w)$ are ill-posed.

So what can we do in order to show that maximum-likelihood decoding of Reed-Solomon codes is computationally hard even with unlimited preprocessing? Here is a sketch of the answer to this question. We can prove that:

> There is an infinite sequence $\mathcal{P}_1, \mathcal{P}_2, \ldots$ of MLD-RSwP$(\cdot, \cdot, \cdot, \cdot)$ problems such that $m_1 \leqslant m_2 \leqslant \cdots$ and $|\mathcal{D}_1| < |\mathcal{D}_2| < \cdots$ with the following property: under a certain assumption that is widely believed to be true, there does not exist a constant $c > 0$ such that for all sufficiently large $i$, each problem $\mathcal{P}_i$ can be solved in time and space at most $(m_i + |\mathcal{D}_i|)^c$. $(\star)$

The precise meaning of "$\mathcal{P}_i$ can be solved in time and space at most $(m_i + |\mathcal{D}_i|)^c$" in $(\star)$ is that there exists a circuit $C_i$ of size at most $(m_i + |\mathcal{D}_i|)^c$ that solves $\mathcal{P}_i$ for every possible input $\underline{y} \in \mathbb{F}_q^n$, where $q = 2^{m_i}$ and $n = |\mathcal{D}_i|$. Observe that we allow different circuits for different problems — that is, the circuit $C_i$ solving $\mathcal{P}_i = $ MLD-RSwP$(m_i, \mathcal{D}_i, k_i, w_i)$ may depend on $m_i$, $\mathcal{D}_i$, $k_i$, and $w_i$. This corresponds to the "nonuniform" version of the class P of polynomial-time decidable languages, where one can use different programs for inputs of different sizes. The resulting complexity class is usually denoted as P/poly. Thus the "assumption that is widely believed to be true" in $(\star)$ is that NP $\not\subseteq$ P/poly or, in words, that not every language in NP has a polynomial-size circuit. It is indeed widely believed that NP $\not\subseteq$ P/poly. In fact, it was shown by Karp and Lipton [14] that if NP $\subseteq$ P/poly then the polynomial hierarchy collapses at the second level, namely $\cup_{i=1}^{\infty} \Sigma_i^p = \Sigma_2^p$. For more details on this and more rigorous definitions of the terms used in this paragraph, we refer the reader to Bruck-Naor [5] and to Papadimitriou [17].

How can one prove a statement such as $(\star)$? The usual way (cf. [5, 8, 16]) to do this is as follows. Start with an NP-complete problem $\Pi$. Then devise a deterministic procedure that converts every instance $\mathcal{I}$ of $\Pi$ into $m, \mathcal{D}, k, w$, and $\underline{y}$ with the following properties:

**P1.** The parameters $m, \mathcal{D}, k, w$ depend only on $\text{size}(\mathcal{I})$, the size of the instance $\mathcal{I}$, and are constructed in time and space that are polynomial in $\text{size}(\mathcal{I})$.

**P2.** The target vector $\underline{y}$ is also constructed in time and space that are polynomial in $\text{size}(\mathcal{I})$, but may depend on the instance $\mathcal{I}$ itself rather than only on its size.

**P3.** The target $\{\underline{y}\}$ is a YES instance of the constructed $\mathsf{MLD\text{-}RSwP}(m, \mathcal{D}, k, w)$ problem if and only if $\mathcal{I}$ is a YES instance of $\Pi$.

For an explanation of this method and for precise definition of $\text{size}(\mathcal{I})$, we again refer the reader to [5, 17]. Here, we take $\Pi$ to be the THREE-DIMENSIONAL MATCHING problem introduced in the previous section. In this case, we can assume, as in [16], that the size of an instance $\{t, \mathcal{T}\}$ of THREE-DIMENSIONAL MATCHING is simply $t$.

The following deterministic procedure combines the ideas of the previous section with a suitably modified version of a reduction due to Lobstein [16]. Incidentally, Lobstein's reduction [16] is by far the simplest way known (to us) to prove that MLD-Linear remains hard with unlimited preprocessing (cf. [5, 8, 19]). Given an instance $\{t, \mathcal{T}\}$ of THREE-DIMENSIONAL MATCHING, we proceed as follows.

**A. COMPUTING THE INTEGER PARAMETERS:** Set $m = 3(t^3 + t)$, $w = t^3 + t$, and $k = 3t^3 - (t+1)$. Let $n = 4t^3$.

**B. COMPUTING THE EVALUATION SET:** As in the previous section, let $q = 2^m$ and construct the finite field $\mathbb{F}_q$. Let $\alpha$ be an arbitrary primitive element of $\mathbb{F}_q$ and fix a basis $\{1, \alpha, \alpha^2, \ldots, \alpha^{m-1}\}$ for $\mathbb{F}_q$ over $\mathbb{F}_2$. Let $\mathcal{U} = \{1, 2, \ldots, t\}$, and impose an arbitrary order on the $t^3$ triples in $\mathcal{U} \times \mathcal{U} \times \mathcal{U}$, say $(a_1, b_1, c_1)$, $(a_2, b_2, c_2), \ldots, (a_{t^3}, b_{t^3}, c_{t^3})$. Define*

$$x_j = \begin{cases} \alpha^{a_j-1} + \alpha^{t+b_j-1} + \alpha^{2t+c_j-1} + \alpha^{3t+j-1} & \text{for } 1 \leqslant j \leqslant t^3 \\ \alpha^{3t+(j-t^3)-1} + \alpha^{3t+j-1} + \alpha^{3t+(j+t^3)-1} & \text{for } t^3 < j \leqslant 2t^3 \\ \alpha^{3t+(j-t^3)-1} + \alpha^{3t+j-1} & \text{for } 2t^3 < j \leqslant 3t^3 \\ \alpha^{3t+(j-t^3)-1} & \text{for } 3t^3 < j \leqslant 4t^3 \end{cases} \quad (16)$$

This produces $n = 4t^3$ distinct nonzero elements $x_1, x_2, \ldots, x_n \in \mathbb{F}_q$. Set the evaluation set $\mathcal{D}$ to $\{x_1, x_2, \ldots, x_n\}$.

---

*The evaluation points $x_1, x_2, \ldots, x_n$ may be better understood in terms of the matrix $W$, defined in (19), whose columns are binary representations of $x_1, x_2, \ldots, x_n$ with respect to the basis $\{1, \alpha, \alpha^2, \ldots, \alpha^{m-1}\}$.

**C. COMPUTING THE TARGET VECTOR:** Let $\underline{\chi}_{\mathcal{T}} = (\chi_1, \chi_2, \ldots, \chi_{t^3})$ be the characteristic vector of $\mathcal{T} \subseteq \mathcal{U} \times \mathcal{U} \times \mathcal{U}$. That is $\chi_j = 1$ if the $j$-th triple $(a_j, b_j, c_j)$ of $\mathcal{U} \times \mathcal{U} \times \mathcal{U}$ belongs to $\mathcal{T}$, and $\chi_j = 0$ otherwise. Compute

$$\gamma \stackrel{\text{def}}{=} \sum_{j=1}^{3t} \alpha^{j-1} + \alpha^{3t-1}\left(\alpha^{t^3}+1\right) \sum_{j=1}^{t^3} \chi_j \alpha^j + \alpha^{2t^3+3t-1} \sum_{j=1}^{t^3} \alpha^j \quad (17)$$

Thus $\gamma$ is an element of $\mathbb{F}_q$ that corresponds to the binary $m$-tuple $(\mathbf{1}, \underline{\chi}_{\mathcal{T}}, \underline{\chi}_{\mathcal{T}}, \mathbf{1})$ in the chosen basis, where the first $\mathbf{1}$ in $(\mathbf{1}, \underline{\chi}_{\mathcal{T}}, \underline{\chi}_{\mathcal{T}}, \mathbf{1})$ is the all-one vector of length $3t$ while the second $\mathbf{1}$ is the all-one vector of length $t^3$. From here, proceed exactly as in the previous section: for each $j = 1, 2, \ldots, w+1$, compute

$$z_j \stackrel{\text{def}}{=} \frac{\gamma - \sum_{\substack{i=1 \\ i \neq j}}^{w+1} x_i}{\prod_{\substack{i=1 \\ i \neq j}}^{w+1} (x_j - x_i)} \quad \text{and} \quad \varphi_j \stackrel{\text{def}}{=} \prod_{\beta \in \mathbb{F}_q \backslash \mathcal{D}} (x_j - \beta) \quad (18)$$

Set the target vector $\underline{y}$ to $(z_1/\varphi_1, z_2/\varphi_2, \ldots, z_{w+1}/\varphi_{w+1}, 0, 0, \ldots, 0)$. In other words: $y_j = z_j/\varphi_j$ for $j = 1, 2, \ldots, w+1$, and $y_j = 0$ otherwise.

We will refer to the foregoing computation as the 3-DM/MLD-RSwP *conversion procedure*. It should be evident from Lemma 1 that this procedure runs in time and space that are polynomial in $t$. Furthermore, it is clear that $m$, $k$, $w$ in Step A and $\mathcal{D}$ in Step B depend only on $t$. Thus properties **P1** and **P2** above are satisfied, and it remains to prove property **P3**.

To this end, consider the $m \times n$ (or $3(t^3 + t) \times 4t^3$) binary matrix $W$ having the binary representations of $x_1, x_2, \ldots, x_n$ as its columns. By construction — compare with the definition $x_1, x_2, \ldots, x_n$ in (16) — this matrix has the following structure:

$$W = \begin{bmatrix} U & 0 & 0 & 0 \\ \hline I & I & 0 & 0 \\ \hline 0 & I & I & 0 \\ \hline 0 & I & I & I \end{bmatrix} \quad (19)$$

where $I$ is the $t^3 \times t^3$ identity matrix and $U$ is the $3t \times t^3$ matrix consisting of the binary representations of the $t^3$ triples in $\mathcal{U} \times \mathcal{U} \times \mathcal{U}$ — that is, the $j$-th column of $U$ is the binary representation of $\alpha^{a_j-1} + \alpha^{t+b_j-1} + \alpha^{2t+c_j-1}$ where $(a_j, b_j, c_j)$ is the $j$-th triple in $\mathcal{U} \times \mathcal{U} \times \mathcal{U}$.

**Lemma 6.** *The set $\mathcal{T}$ has a matching if and only if there is a set of exactly $w = t^3 + t$ columns of $W$ that add to the vector $(\mathbf{1}, \underline{\chi}_{\mathcal{T}}, \underline{\chi}_{\mathcal{T}}, \mathbf{1})^t$, which is the binary representation of $\gamma$.*

*Proof.* Since the order imposed on the triples of $\mathcal{U} \times \mathcal{U} \times \mathcal{U}$ is arbitrary, we may assume w.l.o.g. that the triples in $\mathcal{T}$ correspond to the first $|\mathcal{T}|$ columns of the matrix $U$. ($\Rightarrow$) Sup-pose there is a vector $\underline{v} \in \mathbb{F}_2^n$ with $\text{wt}(v) = t^3 + t$ such that $W\underline{v}^t = (\mathbf{1}, \underline{\chi}_\mathcal{T}, \underline{\chi}_\mathcal{T}, \mathbf{1})^t$. Write $\underline{v} = (\underline{v}_1, \underline{v}_2, \underline{v}_3, \underline{v}_4)$, where $\underline{v}_1, \underline{v}_2, \underline{v}_3, \underline{v}_4$ are vectors of length $t^3$. For $i = 1, 2, 3, 4$, let

$$\eta_i \stackrel{\text{def}}{=} \text{the weight of the first } |\mathcal{T}| \text{ positions of } \underline{v}_i$$

$$\bar{\eta}_i \stackrel{\text{def}}{=} \text{the weight of the last } t^3 - |\mathcal{T}| \text{ positions of } \underline{v}_i$$

The structure of the matrix $W$ in (19) along with the fact that $W\underline{v}^t = (\mathbf{1}, \underline{\chi}_\mathcal{T}, \underline{\chi}_\mathcal{T}, \mathbf{1})^t$ imply the following relationships

$$\eta_2 = |\mathcal{T}| - \eta_1 \text{ and } \bar{\eta}_2 = \bar{\eta}_1 \quad (\text{since } \underline{v}_1 + \underline{v}_2 = \underline{\chi}_\mathcal{T}) \tag{20}$$

$$\eta_3 = |\mathcal{T}| - \eta_2 \text{ and } \bar{\eta}_3 = \bar{\eta}_2 \quad (\text{since } \underline{v}_2 + \underline{v}_3 = \underline{\chi}_\mathcal{T}) \tag{21}$$

$$\bar{\eta}_4 = t^3 - |\mathcal{T}| \text{ and } \eta_4 = 0 \quad (\text{since } \underline{v}_4 = \mathbf{1} - (\underline{v}_2 + \underline{v}_3) = \mathbf{1} - \underline{\chi}_\mathcal{T}) \tag{22}$$

among $\eta_1, \eta_2, \eta_3, \eta_4$ and $\bar{\eta}_1, \bar{\eta}_2, \bar{\eta}_3, \bar{\eta}_4$. Using (20), (21), (22) in conjunction with the fact that $\text{wt}(\underline{v}) = \eta_1 + \eta_2 + \eta_3 + \eta_4 + \bar{\eta}_1 + \bar{\eta}_2 + \bar{\eta}_3 + \bar{\eta}_4 = t^3 + t$, we obtain

$$\eta_1 + 3\bar{\eta}_1 = t \tag{23}$$

But $\text{wt}(\underline{v}_1) = \eta_1 + \bar{\eta}_1 \geqslant t$, since $U\underline{v}_1^t = \mathbf{1}^t$ and the weight of each column of $U$ is 3. In conjunction with (23), this implies that

$$\eta_1 = t \quad \text{and} \quad \bar{\eta}_1 = 0$$

This means that there are some $t$ columns among the first $|\mathcal{T}|$ columns of $U$ (corresponding to the triples in $\mathcal{T}$) that add (mod 2) to the all-one vector. Hence, there is a matching in $\mathcal{T}$.
($\Leftarrow$) Conversely, suppose there is a matching in $\mathcal{T}$. We then take $\underline{v}_1$ to be the binary vector of length $t^3$ and weight $t$ whose nonzero positions are given by the corresponding $t$ columns of $U$. Setting

$$\underline{v}_2 = \underline{\chi}_\mathcal{T} - \underline{v}_1, \quad \underline{v}_3 = \underline{v}_1, \quad \text{and} \quad \underline{v}_4 = \mathbf{1} - \underline{\chi}_\mathcal{T} \tag{24}$$

it is easy to verify that the vector $\underline{v} = (\underline{v}_1, \underline{v}_2, \underline{v}_3, \underline{v}_4)$ satisfies $W\underline{v}^t = (\mathbf{1}, \underline{\chi}_\mathcal{T}, \underline{\chi}_\mathcal{T}, \mathbf{1})^t$ and has weight $t + (|\mathcal{T}| - t) + t + (t^3 - |\mathcal{T}|) = t^3 + t$. ∎

To prove that 3-DM/MLD-RSwP conversion procedure satisfies property **P3**, it remains to combine Lemma 6 with the results of the previous section.

**Lemma 7.** *The set $\mathcal{T}$ has a matching if and only if there is a codeword $\underline{c} \in \mathbb{C}_q(\mathcal{D}, k)$ such that $d(\underline{c}, \underline{y}) \leqslant w$, where $q = 2^m$ and $m, k, \mathcal{D}, w, \underline{y}$ are the values computed from $\{t, \mathcal{T}\}$ in the 3-$\bar{\text{D}}$M/MLD-RSwP conversion procedure.*

*Proof.* Let $H$ be the $(w+1) \times n$ parity-check matrix in (6), but with $x_1, x_2, \ldots, x_n$ now defined by (16). Using Lemma 6 and proceeding exactly as in Lemma 2, we conclude that $\mathcal{T}$ has a matching if and only if there is a vector $\underline{v} \in \mathbb{F}_q^n$ of weight $\leqslant w$ such that

$$H\underline{v}^t = (0, 0, \ldots, 0, 1, \gamma)^t$$

where $\gamma$ is given by (17). By Corollary 3 and Lemma 4 of the previous section, this happens if and only if there is a codeword $\underline{c} \in \mathbb{C}_q(\mathcal{D}, k)$ such that $d(\underline{c}, \underline{y}) \leqslant w$. ∎

**Theorem 8.** *There is an infinite sequence of Reed-Solomon codes $\{\mathbb{C}_{2^{m_i}}(\mathcal{D}_i, k_i)\}_{i \geqslant 1}$, that can be explicitly specified in terms of the underlying fields $\mathbb{F}_{2^{m_1}}, \mathbb{F}_{2^{m_2}}, \ldots$, evaluation sets $\mathcal{D}_1 \subseteq \mathbb{F}_{2^{m_1}}, \mathcal{D}_2 \subseteq \mathbb{F}_{2^{m_2}}, \ldots$, and dimensions $k_1, k_2, \ldots$, such that the following holds: unless $\mathrm{NP} \subseteq \mathrm{P}/\mathrm{poly}$ and the polynomial hierarchy collapses at the second level, there is no polynomial-size family of circuits $\{C_i\}_{i \geqslant 1}$ so that $C_i$ solves the maximum-likelihood decoding problem for the code $\mathbb{C}_{2^{m_i}}(\mathcal{D}_i, k_i)$, for all $i = 1, 2, \ldots$.*

*Proof.* Lemma 7 proves that the 3-DM/MLD-RSwP conversion procedure satisfies properties **P1**, **P2**, and **P3**. Since THREE-DIMENSIONAL MATCHING is NP-complete, this immediately implies the theorem (see the discussion at the beginning of this section). ∎

Theorem 8 is our main result in this section. In plain language, this theorem says that there exist Reed-Solomon codes for which maximum-likelihood decoding is computationally hard even if unlimited preprocessing of the code is allowed.

## 4. Discussion and open problems

We begin this section with a disclaimer, which also leads to an interesting open problem. The 3-DM/MLD-RS conversion procedure of Section 2 produces a specific class of Reed-Solomon codes, and Theorem 5 says that there exist codes in this class that are hard to decode (unless $\mathrm{P} = \mathrm{NP}$). However, since $|\mathcal{D}| = |\mathcal{T}| \leqslant t^3$ while $|\mathbb{F}_{2^m}| = 2^{3t}$ in our conversion procedure, all the codes in this class use only a tiny fraction of the underlying field as evaluation points. Thus our hardness results do not apply if, say, all the field elements (or all the nonzero field elements) are taken as evaluation points, as is often the case with Reed-Solomon codes. On the other hand, the algebraic decoding algorithms for Reed-Solomon codes [3, 12, 21, 24] do not take advantage of this fact and work just as well for arbitrary sets of evaluation points (such as those produced by our conversion procedure).

Nevertheless, it remains an intriguing open question whether a similar hardness result can be established for Reed-Solomon codes that use the entire field (or a large part thereof) as their set of evaluation points. The proof of this (if it exists) will probably require new techniques, and might also pave the way for establishing NP-hardness of maximum-likelihood decoding for primitive binary BCH codes. We observe that such a proof would immediately imply hardness with unlimited preprocessing, since in this situation the code is essentially fixed: only its rate and the received syndrome are part of the input.

---

We next record a simple corollary to our main result. It is well known [6, Chapter 10, p. 281] that the covering radius $\rho$ of an $(n, k)$ Reed-Solomon code $\mathbb{C}_q(\mathcal{D}, k)$ is given by $\rho = n - k$. A vector $\underline{y} \in \mathbb{F}_q^n$ is said to be a ***deep hole*** of $\mathbb{C}_q(\mathcal{D}, k)$ if the distance from $\underline{y}$ to (the closest codeword of) this code is exactly $\rho$. We observe that the value of $w$ in the reduc-tion of Section 2 is $n - k - 1 = \rho - 1$, so that we are asking whether there exists a codeword $\underline{c} \in \mathbb{C}_q(\mathcal{D}, k)$ such that $d(\underline{c}, \underline{y}) \leqslant \rho - 1$. This is equivalent to the question: is $\underline{y}$ a deep hole of $\mathbb{C}_q(\mathcal{D}, k)$? Hence, Theorem 5 immediately implies the following result.

**Corollary 9.** *It is NP-hard to determine whether a given vector $\underline{y} \in \mathbb{F}_q^n$ is a deep hole of a given Reed-Solomon code $\mathbb{C}_q(\mathcal{D}, k)$.*

In fact, it is easy to see from the proof of Lemma 2 that the distance from the vector $\underline{y}$ constructed in the 3-DM/MLD-RS conversion procedure to $\mathbb{C}_q(\mathcal{D}, k)$ is *at least $w = \rho - 1$*. Thus an even more specialized task is NP-hard: given a vector which is either at distance $\rho$ or at distance $\rho - 1$ from $\mathbb{C}_q(\mathcal{D}, k)$, determine which is the case. Note that the reduction in Section 3 still has the property that $w = n - k - 1 = \rho - 1$. Thus identifying deep holes of a Reed-Solomon code (or deciding whether a given vector is at distance $\rho$ or $\rho - 1$ from the code) is computationally hard even if unlimited preprocessing of the code is allowed.

Concerning the results of Section 3, we observe that a polynomial-time maximum-likelihood decoding algorithm for some specific Reed-Solomon codes (if it exists) *must* make essential use of the structure of the evaluation sets for these codes. Section 3 shows that, assuming NP does not have polynomial-size circuits, there is no generic representation of the evaluation points that would permit polynomial-time maximum-likelihood decoding.

---

We conclude the paper with two more open problems. First, it would be interesting to establish NP-hardness of maximum-likelihood decoding for a nontrivial family of *binary* codes. Straightforward concatenation of Reed-Solomon codes over $\mathbb{F}_{2^m}$ with $(2^m - 1, m, 2^{m-1})$ simplex (Hadamard) codes does not work, since the length of the concatenated code would be exponential in the length of the Reed-Solomon code for our reduction.

Another important open problem is this. As discussed in Corollary 9, maximum-likelihood decoding of Reed-Solomon codes becomes hard when the number of errors is large — one less than the covering radius of the code. It is an extremely interesting problem to show hardness of *bounded-distance* decoding of Reed-Solomon codes for a smaller decoding radius. At present, there remains a large gap between our hardness results and the decoding radius up to which polynomial-time decoding algorithms are known [12, 15].

**Acknowledgment.** The results reported in Section 2 were derived while both authors were attending the Kodierungstheorie Workshop at the Mathematisches Forschungsinstitut in Oberwolfach, Germany, in December 2003. We would like to use this opportunity to thank the Mathematisches Forschungsinstitut Oberwolfach for its hospitality.